\documentclass[onecolumn,useAMS,usenatbib]{mn2e}

\usepackage{graphicx}
\usepackage{subfigure}
\usepackage{xcolor}
\usepackage{mathtext,bbm,amsmath,amsfonts,amssymb,indentfirst,syntonly,graphicx}
\usepackage{mathtools}
\usepackage{slashbox}
\usepackage[english]{babel}
\usepackage{calc}
\usepackage{tikz}

\def\bc{\begin{center}}
\def\ec{\end{center}}
\def\be{\begin{eqnarray}}
\def\ee{\end{eqnarray}}

\title[Rotation Curves in the Grumiller's Gravity]{Galaxy Rotation Curves  in the Grumiller's Modified Gravity}
\author[H.-N. Lin, M.-H. Li, X. Li and Z. Chang]
        {Hai-Nan Lin$^{1}$\thanks{E-mail: linhn@ihep.ac.cn (corresponding author);},
         Ming-Hua Li$^{1}$\thanks{E-mail: limh@ihep.ac.cn;},
         Xin Li$^{1,2}$\thanks{E-mail: lixin@ihep.ac.cn;} and
         Zhe Chang$^{1,2}$\thanks{E-mail: changz@ihep.ac.cn}\\
$^{1}$Institute of High Energy Physics, Chinese Academy of Sciences, 100049 Beijing, China\\
$^{2}$Theoretical Physics Center for Science Facilities, Chinese Academy of Sciences, 100049 Beijing, China}
\begin{document}

\date{Accepted xxxx; Received xxxx; in original form xxxx}

\pagerange{\pageref{firstpage}--\pageref{lastpage}} \pubyear{2012}

\maketitle

\label{firstpage}

\begin{abstract}
The effective potential in the Grumiller's modified gravity [D. Grumiller, Phys. Rev. Lett {\bf 105}, 211303 (2010)] includes the Newtonian potential and a Rindler term. The fitting to the rotation curve data of eight galaxies suggests a universal Rindler acceleration $a\approx 0.30\times 10^{-10}$ m s$^{-2}$. We do a two-parameter fit first, with the mass-to-light ratio ($\Upsilon_*$) and the Rindler acceleration ($a$) as free parameters. It is found  that the data of six out of the eight galaxies fit well with the prediction of theory in the range $0\lesssim r \lesssim 40$ kpc, although the theoretical curves show a tendency of arising beyond this range.  The Rindler accelerations of the six well-fitted galaxies have the same magnitude, with an average value $\bar{a}\approx0.30\times 10^{-10}$ m s$^{-2}$. Inspired by this fact, we then carry out a one-parameter ($\Upsilon_*$) fit to the six galaxies, with $a$ fixed at $\bar{a}$, and find that the theory can still reproduce the observation. The value of Rindler acceleration we get here is a quarter of that of Milgrom's MOND. For the rest two galaxies, NGC5055 and DDO154, the fitting results are significantly improved if the photometric scale length ($h$) is included as another free parameter.
\end{abstract}


\section{Introduction}
The observations on rotation velocities of stars around a galaxy center show significant discrepancies from Newtonian theory. According to Newtonian gravity, the rotation velocity is inversely proportional to the square root of distance from the center of a galaxy. However, the observed data often show an asymptotically flat rotation curve out to the furthest data points. Generally, there are three different ways to solve this problem. The most direct assumption is that there are a large number of nonluminous matters that have not been measured yet \citep{Begeman1991,Persic1996,Chemin:2011mf}. However, after decades of years heavy research, no direct evidences of the existence of dark matter have been found. This inspires the astronomers and physicists to search for other explanations of the discrepancy between the Newtonian dynamical mass and the luminous mass. The most popular way is to modified the Newton dynamics (MOND) \citep{Milgrom1983a,Milgrom1983b}. With only one universal parameter, ie., the critical acceleration $a_0$, MOND has achieved great success in explaining the mass discrepancy problem \citep{Sanders1996,Sanders:1998gr}.

Besides MOND, it is also possible to  modify the Newtonian gravity (MOG).  According to MOG, Newtonian gravity is invalid at galaxy scales. Such theories include nonsymmetric gravity theory (NGT, \citet{Moffat:1995}), metric-skew-tensor gravity (MSTG, \citet{Moffat:2005,Brownstein:2005zz}), scalar-tensor-vector gravity (STVG, \citet{Moffat:2006}), Carmelian general relativity (CGR, \citet{Carmeli2000}), Horava-Lifshitz gravity theory (H-L theory, \citet{Horava2009a,Horava2009b,Horava2009c,Cardone2010,Cardone2012}), etc.. All of them can in a large degree explain the observed rotation curves.

Two years ago, \citet{Grumiller:2010bz} proposed an effective model for gravity at large distance, and obtained an effective potential in leading order consists of a Newtonian potential ($\propto 1/r$) and a Rindler term ($\propto r$). \citet{Grumiller:2011gg} showed that the Rindler force is  capable of explaining about 10\% of the Pioneer anomaly \citep{Anderson1998}, and simultaneously ameliorates the shape of galactic rotation curves. Most importantly, the Rindler term doesn't spoil the solar-system precision tests.

In this paper, we investigate the galaxy rotation curves in the Grumiller's modified gravity \citep{Grumiller:2010bz}. The rest of the paper is organized as follows: In section \ref{sec:theory}, we review briefly the Grumiller's modified gravity theory. In section \ref{sec:application}, we do a best-fit procedure to constrain the parameters. The two-parameter fit shows that six out of eight galaxies have approximately the same Rindler acceleration $\bar{a}\approx0.30\times 10^{-10}$ m s$^{-2}$. Then we carry out a one-parameter fit, with $a$ fixed at $\bar{a}$. It is found that the theory can still reproduce the observation. For the rest two galaxies, the fitting results are significantly improved if the photometric scale length is included as a free parameter. Finally,  discussions and conclusions are given in section \ref{sec:discussion}.

\section{Grumiller's modified gravity}\label{sec:theory}

In Grumiller's modified gravity theory \citep{Grumiller:2010bz}, the spacetime is described by a spherically symmetric metric
\begin{equation}\label{eq:line-element}
  ds^2=g_{\alpha\beta}dx^{\alpha}dx^{\beta}+\Phi^2(d\theta^2+\sin^2\theta d\phi^2),
\end{equation}
where the two dimensional metric $g_{\alpha\beta}(x^{\gamma})$ and the surface radius $\Phi(x^{\gamma})$ depends only on the coordinates $x^{\gamma}=\{t,r\}$. The Einstein-Hilbert action and the matter action are given by
\begin{equation}\label{eq:EH-action}
  S_{\rm EH}=-\int d^4x\sqrt{-g}R,
\end{equation}
and
\begin{equation}\label{eq:matter-action}
  S_{\rm m}=-\int d^4x\sqrt{-g}\mathcal{L}_{\rm m},
\end{equation}
where $R$ is the Ricci scalar and $\mathcal{L}_{\rm m}$ is the Lagrangian of the matter part.

Generally, a 4-dimensional spherical spacetime manifold $M$ can be decomposed to the direct product of a 2-dimensional radial sub-manifold $L$ and a 2-dimensional angular sub-manifold $S$, $M=L\otimes S$. The ``spherical reduction'' process \citep{Grumiller2001} simplifies the 4-dimensional Einstein-Hilbert action to a specific 2-dimensional dilation gravity model (see Appendix A for details)
\begin{equation}\label{eq:action2}
  S_{\rm EH}=-\int d^2x\sqrt{-g}[f(\Phi)R+2(\partial\Phi)^2-2].
\end{equation}
For the same reason, the matter action can be reduced to a 2-dimensional one
\begin{equation}\label{eq:matter-action2}
  S_{\rm m}=-\int d^2x\sqrt{-g}\Phi^2 \mathcal{L}_{\rm m}.
\end{equation}
Note that the metric $g_{\mu\nu}$ and the Ricci scalar $R$ in Eq.(\ref{eq:action2}) and Eq.(\ref{eq:matter-action2}) belong to the 2-dimensional sub-manifold $L$. For simplicity, we will not explicitly distinguish the symbols on $M$ and $L$ here and after. Since we are only interested in the vacuum solution, we will not discuss the matter part bellow.

If we allow for IR modification to the model but require it to be power-counting renormalizable and non-singularity of curvature for large $\Phi$ , the action Eq.(\ref{eq:action2}) can be generalized to the form \citep{Grumiller:2010bz}
\begin{equation}\label{eq:action3}
  S=-\int d^2x\sqrt{-g}[f(\Phi)R+2(\partial\Phi)^2-2V(\Phi)],
\end{equation}
where
\begin{equation}
V(\Phi)\equiv 3\Lambda\Phi^2-4a\Phi-1.
\end{equation}
Here, $\Lambda$ and $a$ are two constants. We will show later that $\Lambda$ is related to the cosmological constant, and $a$ corresponds to the Rindler acceleration.

Variation of the action in Eq.(\ref{eq:action3}) gives the equations of motion (EOM) \citep{Grumiller:2010bz}
\begin{equation}\label{eq:field-equation}
  \begin{dcases} R=\frac{2}{\Phi}g^{\alpha\beta}\nabla_{\alpha}\partial_{\beta}\Phi+6\Lambda-\frac{4a}{\Phi}\ ,\\
  2\Phi(\nabla_{\mu}\partial_{\nu}-g_{\mu\nu}\nabla^{\alpha}\partial_{\alpha})\Phi-g_{\mu\nu}(\partial\Phi)^2=g_{\mu\nu}V(\Phi)\ ,
  \end{dcases}
\end{equation}
where the first equality is obtained from varying the action with respect to the scalar field $\Phi$, and the second equality is obtained from varying the action with respect to the 2-dimensional metric $g^{\mu\nu}$. `$\nabla_{\mu}$' represents covariant derivatives.

The general solution of Eq.(\ref{eq:field-equation}) can be obtained using the gauge theoretic formulation given by \citet{Cangemi1992}. In the Schwarzschild gauge where the metric is diagonal, we can find the solution of the above EOM \citep{Grumiller:2010bz}
\begin{equation}\label{eq:solution}
  \begin{dcases}
    g_{\alpha\beta}dx^{\alpha}dx^{\beta}=-K^2dt^2+\frac{dr^2}{K^2},\\
    \Phi=r,
  \end{dcases}
\end{equation}
where
\begin{equation}\label{eq:Kr}
  K^2\equiv1-\frac{2M}{r}-\Lambda r^2+2ar.
\end{equation}
If $a=\Lambda=0$, Eq.(\ref{eq:solution}) reduces to the Schwarzschild solution in general relativity. While if $M=\Lambda=0$, Eq.(\ref{eq:solution}) reduces to the 2-dimensional Rindler metric \citep{Wald1984}. Since the 2-dimensional sub-manifold $L$ is embedded into the 4-dimensional manifold $M$, the solution Eq.(\ref{eq:solution}) and Eq.(\ref{eq:Kr}) induces a 4-dimensional metric on $M$,
\begin{equation}\label{eq:metric}
  g_{\mu\nu}={\rm diag}(-K^2,1/K^2,r^2,r^2\sin^2\theta).
\end{equation}
A straightforward calculation gives the the Ricci scalar of the 4-dimensional manifold (see Appendix B for details)
\begin{equation}\label{eq:Ricci-scalar}
  R=-\frac{2K^2\Phi''}{\Phi}+\frac{2}{\Phi^2}(-2KK'\Phi\Phi'-K^2\Phi'^2-K^2\Phi\Phi''+1)\, ,
\end{equation}
where the primes denote derivatives with respect to the radial distance $r$.
Substituting Eq.(\ref{eq:solution}) and Eq.(\ref{eq:Kr}) into Eq.(\ref{eq:Ricci-scalar}), we obtain
\begin{equation}
  R=6\Lambda-\frac{8a}{r}.
\end{equation}

Consider a point particle with energy $E$ and angular momentum $\ell$ moving along a geodesic in the plane $\theta=\pi/2$ in the background of the metric Eq.(\ref{eq:metric}). The conservation of energy and angular momentum gives \citep{Wald1984}
\begin{equation}
  \begin{dcases}\label{eq:conservation}
    K^2\frac{dt}{d\tau}=E, \\
    r^2\frac{d\varphi}{d\tau}=\ell.
  \end{dcases}
\end{equation}
The normalization of 4-velocity ($g_{\mu\nu}u^{\mu}u^{\nu}=-1$) gives another equality
\begin{equation}\label{eq:normalization}
    -K^2\left(\frac{dt}{d\tau}\right)^2+\frac{1}{K^2}\left(\frac{dr}{d\tau}\right)^2+r^2\left(\frac{d\varphi}{d\tau}\right)^2=-1.
\end{equation}
Using Eq.(\ref{eq:conservation}), Eq.(\ref{eq:normalization}) can be transformed into the following form
\begin{equation}\label{eq:conservation2}
    \frac{1}{2}\left(\frac{dr}{d\tau}\right)^2=\frac{1}{2}E^2-V^{\rm eff},
\end{equation}
where
\begin{equation}\label{eq:effective-potential}
V^{\rm eff}\equiv \frac{K^2}{2}\left(1+\frac{\ell^2}{r^2}\right)=-\frac{M}{r}+\frac{\ell^2}{2r^2}-\frac{M\ell^2}{r^3}-\frac{\Lambda r^2}{2}+ar(1+\frac{\ell^2}{r^2})+...
\end{equation}
is the effective potential. We have dropped the constant terms on the right-hand-side(r.h.s.) of Eq.(\ref{eq:effective-potential}).

The first two terms on the r.h.s. of Eq.(\ref{eq:effective-potential}) are the classical Newtonian potential and the centrifugal barrier, respectively. The third term is the general relativity correction.
The fourth term is the cosmological constant term. The last term, which is proportional to the Rindler acceleration $a$ ,
is peculiar to Grumiller's modified gravity model. Since the Rindler acceleration term
\begin{equation}\label{eq:RA-term}
  V^{\rm RA}=ar(1+\frac{\ell^2}{r^2})
\end{equation}
depends on the angular momentum, this may imply that the modified gravity depends on the orbital eccentricities of stars, but just their positions. However, the second term on the r.h.s. of Eq. (\ref{eq:RA-term}) are much small than the first term. In the International System of Units (SI), Eq. (\ref{eq:RA-term}) takes the form
\begin{equation}
V^{\rm RA}=ar(1+\frac{\ell^2}{c^2r^2}) = ar\left(1+\frac{(vr)^2}{c^2r^2}\right)= ar(1+\frac{v^2}{c^2}),
\end{equation}
where $v$ is the rotational velocity of a star moving around the center of the galaxy and $c$ is the speed of light. For most galaxies, $v$ is of the order of magnitude about $10^2$ km s$^{-1}$ and we have $v^2/c^2 \ll 1$. Thus, the second term on the r.h.s. of Eq. (\ref{eq:RA-term}) can be neglected compared to the first term.

In the case of vanishing cosmological constant and vanishing angular momentum (ie., $\Lambda=\ell=0$), the effective force corresponding to Eq.(\ref{eq:effective-potential}) is given as
\begin{equation}
  F^{\rm eff}=-\frac{\partial V^{\rm eff}}{\partial r}\bigg{|}_{\ell=\Lambda=0}=-\frac{M}{r^2}-a.
\end{equation}
Thus, the rotation velocity of a testing particle moving in the potential of a point particle of mass $M$ reads
\begin{equation}\label{eq:Grumiller-velocity}
  v(r)=\sqrt{\frac{M}{r}+ar}=\sqrt{v_N^2+ar},
\end{equation}
where $v_N\equiv\sqrt{M/r}$ is the rotation velocity derived from Newtonian dynamics. This velocity has the asymptotic behavior
\begin{equation}\label{eq:limit-v}
  v(r)=\begin{cases}v_N(r)\quad \ \ & r\rightarrow 0\ ,\\
  \sqrt{ar}\quad \ \ & r\rightarrow \infty\, .
  \end{cases}
\end{equation}
At small distances, the rotation velocity reduces to the Newtonian case. The Rindler force dominates only at sufficiently large distances, such as the galaxy scales.

\section{Best-fit procedure}\label{sec:application}
To calculate the theoretical rotation velocity, we should know the mass distribution of a galaxy. Generally, the mass in a galaxy contains two components: gas (mainly HI and He) and stellar disk. Some early type galaxies may also contain a bulge in the center. However, the mass of the bulge is often much  smaller than that of the stellar disk, and we will neglect it for simplicity.  Since the THINGS (The HI Nearby Galaxy Survey)\footnote{http://www.mpia-hd.mpg.de/THINGS/Data.html \citep{Walter2008}.} brought an unprecedented level of precision to the measurement of the rotation curves of galaxies, we choose the THINGS galaxies as the samples. The profile of neutral hydrogen (HI) is read directly from the THINGS data cube (robust weighted moment-0) using the task \textsc{ellint} of Groningen Image Processing System (GIPSY)\footnote{http://www.astro.rug.nl/$\sim$gipsy/ (maintained by Hans Terlouw).}. Assuming that HI locates in an infinitely thin disk, we calculate the rotation velocity contributed by  HI ($v_{\rm HI}$) directly using the task \textsc{rotmod} of GIPSY.

The profile of stellar disk is derived from the photometric data. The surface brightness of the stellar disk can be fitted well by an exponential law \citep{Vaucouleurs1959}
\begin{equation}\label{eq:brightness}
  I(r)=I(0)e^{-\frac{r}{h}},
\end{equation}
where $h$ is the scale length and $I(0)$ is the surface brightness at the center of galaxy. The integration of Eq.(\ref{eq:brightness}) gives the total luminosity
\begin{equation}
  L=2\pi I(0)h^2.
\end{equation}
Assuming that the mass-to-light ratio ($\Upsilon_*$) is a constant and the disk is infinitely thin, the mass profile of the stellar disk can be written as
\begin{equation}
  \Sigma(r)=\Sigma(0)e^{-\frac{r}{h}},
\end{equation}
where $\Sigma(0)=\Upsilon_{*}I(0)$ is the central surface mass density. The total mass of the stellar disk is given by
\begin{equation}\label{eq:disk-mass}
  M=\int_0^{\infty} 2\pi r \Sigma(r)dr=2\pi\Sigma(0)h^2.
\end{equation}
The mass-to-light ratio is given as
\begin{equation}\label{eq:mass-to-light}
  \Upsilon_{*}=\frac{M}{L}=\frac{\Sigma(0)}{I(0)}.
\end{equation}
The total luminosity $L$ of a galaxy is related to its absolute magnitude $\mathcal{M}$ as
\begin{equation}\label{eq:M-L-relation}
  \mathcal{M}-\mathcal{M}_{\odot}=-2.5{\rm log}_{10}\frac{L}{L_{\odot}},
\end{equation}
where $\mathcal{M}_{\odot}$ and $L_{\odot}$ are the absolute magnitude and total luminosity of the sun, respectively.

The rotation velocity contributed by the exponential disk can be given as \citep{Freeman1970}
\begin{equation}\label{eq:disk-velocity}
  v_{*}(r)=\sqrt{\frac{M}{r}\gamma(r)},
\end{equation}
where
\begin{equation}
  \gamma(r)\equiv\frac{r^3}{2h^3}\left[I_0\left(\frac{r}{2h}\right)K_0\left(\frac{r}{2h}\right)- I_1\left(\frac{r}{2h}\right)K_1\left(\frac{r}{2h}\right)\right].
\label{gamma}
\end{equation}
Here, $I_{n}$ and $K_{n}$ ($n=0,~1$) are the $n$-th order modified Bessel functions of the first and second kind, respectively.

The Newtonian velocity due to the combined contributions of gas and stellar disk is given by
\begin{equation}\label{eq:newton-velocity}
  v_N=\sqrt{\frac{4}{3}v_{\rm HI}^2+v_{*}^2},
\end{equation}
where the factor 4/3 comes from the contribution of both helium (He) and neutral hydrogen (HI). Here we assume that the mass ratio of He and HI is $M_{\rm He}/M_{\rm HI}=1/3$. Any other gases are  negligible compared to HI and He. Combining Eq.(\ref{eq:Grumiller-velocity}) and Eq.(\ref{eq:newton-velocity}), we get the theoretical rotation velocity in the Grumiller's modified gravity
\begin{equation}\label{eq:fit-formula}
  v(r)=\sqrt{\frac{4}{3}v_{\rm HI}^2(r)+v_{*}^2(r)+ar}.
\end{equation}

In principle, Eq.(\ref{eq:fit-formula}) is only valid to spherically symmetric systems. Since the potential in Eq.(\ref{eq:effective-potential}) is nonlinear to mass, superposition principle as well as Gauss theorem no longer holds, and we can't do an integration over the volume of mass as is done in the case of Newtonian theory to obtain the potential of any mass distribution. Any modified gravity are confronted with this problem. We adopt an approximation as is done in MOND and other modified gravities, by simply extrapolating Eq.(\ref{eq:fit-formula}) to any mass distribution, at least to the axisymmetric cases.

\begin{figure}
  \centering
 \includegraphics[width=16 cm,height=17.8cm]{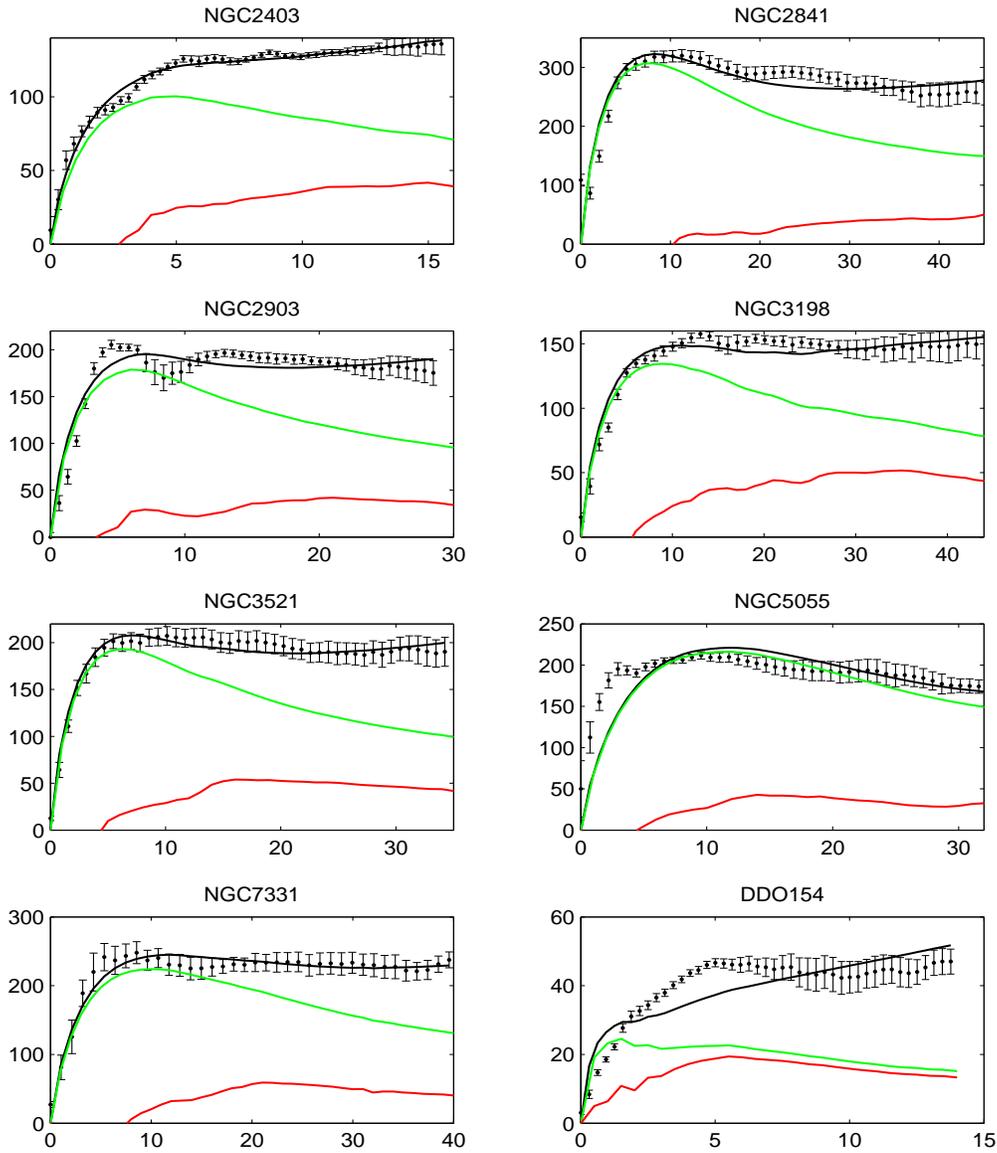}
  \caption{\small{Two-parameter fit (black curves) to the rotation curves of sample galaxies. The $x$-axis is the distance in kpc, and the $y$-axis is the rotation velocity in km~s$^{-1}$. The red curves are the contribution of gas (HI and He). The green curves are the contribution of gas and stellar disk in Newtonian dynamics.}}
  \label{fig:figure1}
\end{figure}

Now we can use Eq.(\ref{eq:fit-formula}) to fit the observed rotation curves. The free parameters are  $\Sigma_0$ (or $\Upsilon_{*}$) and $a$. The sample galaxies and their properties are listed in Table \ref{tab:samples}. The rotation velocity data are read directly from the THINGS data cube (robust weighted moment-1) using the task \textsc{rotcur} of GIPSY, with PA, INCL, $V_{\rm sys}$ and galaxy center (RA and DEC) fixed at the values given in Table \ref{tab:samples}. In order to constrain the parameters, we define the Chi-square as
\begin{equation}\label{eq:chi-aquare}
  \chi^2=\sum_{i=1}^n \frac{[v_i^{\rm obs}-v^{\rm th}(r_i)]^2}{\sigma_i^2},
\end{equation}
where $v_i^{\rm obs}$ is the observed rotation velocity, $v^{\rm th}(r_i)$ is the theoretical velocity  at radius $r_i$ calculated from Eq.(\ref{eq:fit-formula}), and $\sigma_i$ is the uncertainty  of $v_i^{\rm obs}$. Then we employ the least-square method to minimize Eq.(\ref{eq:chi-aquare}). The best-fit results are presented in Figure \ref{fig:figure1}, and the values of parameters are listed in Table \ref{tab:parameters1}.

\begin{figure}
  \centering
 \includegraphics[width=16 cm,height=13.6cm]{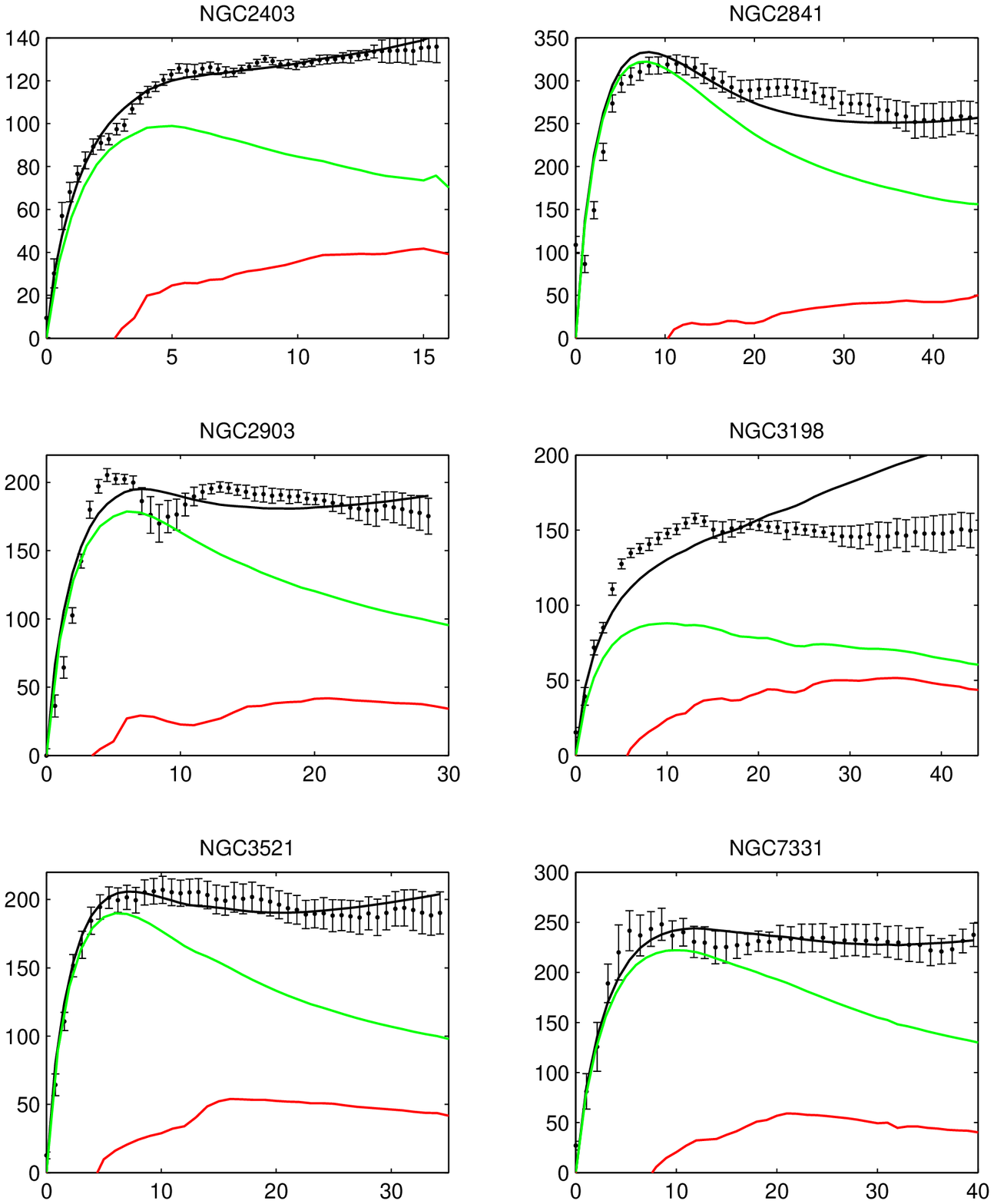}
  \caption{\small{One-parameter fit with $a$ fixed at $\bar{a}\approx0.30\times 10^{-10}{\rm m~s}^{-2}$. The $x$-axis is the distance in kpc, and the $y$-axis is the rotation velocity in km~s$^{-1}$. The red curves are the contribution of gas (HI and He). The green curves are the contribution of gas and stellar disk in Newtonian dynamics.}}
  \label{fig:figure2}
\end{figure}

\begin{figure}
  \centering
 \includegraphics[width=16 cm,height=3.6cm]{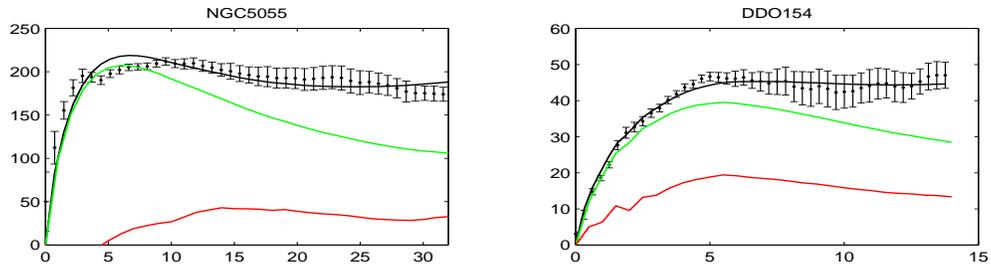}
  \caption{\small{Three-parameter fit with $\Sigma_0$, $a$ and $h$ free. The $x$-axis is the distance in kpc, and the $y$-axis is the rotation velocity in km~s$^{-1}$. The red curves are the contribution of gas (HI and He). The green curves are the contribution of gas and stellar disk in Newtonian dynamics.}}
  \label{fig:figure3}
\end{figure}

From Table \ref{tab:parameters1}, we can see that the Rindler acceleration $a$ has the same magnitude for all the eight galaxies except for NGC5055 and DDO154, with an average value $\bar{a}\approx0.30\times 10^{-10} {\rm m~s}^{-2}$. Inspired by this fact, we carry out a one-parameter fit procedure, with $a$ fixed at $\bar{a}$ and $\Sigma(0)$ as the only free parameter. The best-fit results are shown in Figure \ref{fig:figure2}, and the values of parameters are listed in Table \ref{tab:parameters2}. From Figure \ref{fig:figure2}, we find that the theory can still fits the observations (except for NGC3198), although the theoretical curves show a tendency of arising beyond the data range. For the two poorly-fitted galaxies, NGC5055 and DDO154, we include the disk scale length $h$ as a free parameter and do a three-parameter fit. The results are plotted in Figure \ref{fig:figure3}, and the best-fit parameters are listed in Table \ref{tab:parameters3}. It turns out that the fitting results are significantly improved. However, the best-fit scale lengths far deviate from the photometric scale lengths.  We don't know if it mean that the mass scale length differs from that of photometry.

\section{Discussion and conclusion}\label{sec:discussion}

In this paper, we have done a best fit procedure to the rotation curves of eight THINGS galaxies in the framework of Grumiller's modified gravity. We did a two-parameter fit first, with $\Sigma_0$ and $a$ as free parameters. For the eight sample galaxies, six of them fit well in the range $0\lesssim r\lesssim 40$ kpc. We found that the Rindler accelerations of each galaxies share the same magnitude, with an average value $\bar{a}\approx 0.30\times 10^{-10}$ m s$^{-2}$. The Rindler acceleration we got here is a quarter of the critical acceleration of Milgrom's MOND ($a_0\approx 1.2\times 10^{-10}$ m~s$^{-2}$).  Inspired by this fact, we assumed that $a$ may be a universal constant, and did another one-parameter fit to the six well-fitted galaxies, with $a$ fixed at $\bar{a}$ and $\Upsilon_*$ as the only free parameter. It turned out that the theoretical rotation curves can still reproduce the observed data, with only one exception. The theoretical velocity of NGC3198 shows a tendency of sharply arising after $\sim 20$ kpc. For the rest two galaxies, NGC5055 and DDO154, if we include the disk scale length as a free parameter, the fitting results are significantly improved. Thus, the Rindler acceleration of all the eight galaxies, except for DDO154, can be unified to $\bar{a}$. However, the Rindler acceleration of DDO154, which is a gas rich dwarf galaxy, is one order of magnitude smaller.

A question may arise on applying a spherical symmetry built-in theory to the disk galaxies. In fact, spherical solutions are often obtained for theoretical studies of a gravity model or theory. The solution is then extrapolated to other asymmetric systems under certain conditions. A good example is Bekenstein's TeVeS theory. In \cite{Bekenstein2005}, the famous MOND was obtained in a spherically symmetric situation. He also pointed out that for a generically asymmetric system, a curl term $\nabla \times \textbf{\textit{h}}$ arises in the solution of the Poisson's equation of TeVeS ($\textbf{\textit{h}}$ is some regular vector field). This term falls off rapidly at large distances thus is negligible well outside the system. The theory applies well outside any non-spherical galaxy just as it applies anywhere inside a spherical one. In his paper, Bekenstein showed that it is reasonable to use a spherical symmetry built-in theory to study a non-spherical matter system.

The above conclusion is not valid under some circumstances, such as the interior and the near exterior of a non-spherical galaxy. There also exist qualitative differences: stars cannot escape in a spherical MOND potential, but are able to escape when a non-spherical external field is included \citep{Wu2008}. In these situations, one should solve the Poisson's equation for an asymmetric matter distribution by numerical methods like \cite{Milgrom1986}. For MOND, this method has been used to give fruitful results in regard to low surface brightness disk galaxies, dwarf spheroidal galaxies, and the outer regions of spiral galaxies \citep{deBlok1998,Llinares2008,Swaters2010,Angus2012}. Therefore, to apply Grumiller's spherically symmetric model to calculate the rotation curves of non-spherical spiral galaxies, one should follow Milgrom's way to solve numerically the Poisson's equation of Grumiller's theory for axisymmetric matter distributions, e.g. an exponential stellar disk $\rho(r,z)$ (in cylindrical coordinates $r,\theta,z$). Like MOND, this procedure is computationally expensive. Another way was given by \cite{Swaters2010}. For MOND, they used the usual formula $\mathbf{g}_\mathrm{M}=\mathbf{g}_{\mathrm{N}}/\mu(\mathbf{g}_\mathrm{M}/a_0)$ ($a_0$ is the MOND acceleration parameter and $\mu(x)\equiv x/\sqrt{1+x^2}$) to obtain the MOND acceleration $\mathbf{g}_\mathrm{M}$ as a function of the Newtonian acceleration $\mathbf{g}_\mathrm{N}$, i.e.
\begin{equation}\label{eq:g in gN}
\mathbf{g}_\mathrm{M}^2=\mathbf{g}_\mathrm{N}^2\big{(}1+\sqrt{1+4a_0^2/\mathbf{g}_\mathrm{N}^2}\,\big{)}/2\,.
\end{equation}
Using the above relation, the circular velocity $v_{\rm M}(r)$ within the MOND framework can be expressed as a function of $a_0$ and the Newtonian baryonic contribution $v_{\rm N}(r)$ as $v_{\rm M}^2 (r)=v_{\rm N}^2(r)\big{[} \frac{1}{2}\big{(}1+ \sqrt{1+\left(2ra_0/v_{\rm N}^2(r)\right)^2}\,\big{)}\big{]}^{1/2}$, where $v_{\rm N}(r)\equiv\sqrt{G_\mathrm{N} M/r}$, $G_\mathrm{N}$ is the Newtonian gravity constant. The stellar contribution of $v_{\rm N}(r)$ was obtained by solving the Newtonian Poisson's equation for an axisymmetric density distribution $\rho(r,z)\propto e^{-z/h}$ (in cylindrical coordinates ($r,\theta,z$), $h$ is the scale height of the stellar disk). The result was given analytically as Eq.(\ref{eq:disk-velocity}) and (\ref{gamma}) \citep{Freeman1970,Casertano1983}. According to \cite{Milgrom1986}, the differences between the results obtained by using this method and those by numerically solving the Poisson's equation of MOND are usually much smaller than $5\%$, although in some cases it may be much larger \citep{Angus2006}. The second method is computationally more economical so that it has been widely used to calculate the circular velocity for low surface brightness disk galaxies, dwarf and other non-spherical galaxies \citep{Gentile2008,Swaters2010}.

Thus, considering all these mentioned above, it is reasonable to apply Grumiller's gravity model to non-spherical disk galaxies, even though the theory was first contrived in spherical symmetry. All one has to do is to make the Newtonian approximation of the theory and then solve the corresponding Poisson's equation for non-spherical density distributions, like the MOND and other modified gravity theories. In this paper, we took the second path in the above discussions by substituting Eq.(\ref{eq:disk-velocity}) and Eq.(\ref{gamma}) into the relation between the acceleration for Grumiller's model $\mathbf{g}_\mathrm{G}$ and the Newtonian acceleration $\mathbf{g}_\mathrm{N}$, i.e.
\begin{equation}
\mathbf{g}_\mathrm{G}=\mathbf{g}_{\textmd{N}}+a,
\end{equation}
where $a$ is called the Rindler acceleration. This relation has its analog in MOND as Eq.(\ref{eq:g in gN}). Thus, the rotational velocity in Grumiller's modified gravity is given as $v_\mathrm{G}(r)=\sqrt{G_\mathrm{N}M/r+ar}=\sqrt{v_\mathrm{N}^2+ar}$. It is simply the Eq.(\ref{eq:fit-formula}). We based our numerical analysis on this equation.

In addition, a few comments should be given on the parameter $a$. In our numerical study, we regarded the Rindler acceleration as a constant for a certain galaxy. \citet{Grumiller:2010bz} argued that $a$ may vary from system to system. Further more, $a$ is not necessary to be a constant even in the same system: it  may be a function of $r$. For instance, $a$ is one order of magnitude larger in the Sun-Pioneer system than that of in the Galaxy-Sun system. In the Earth-Satellite system, it may be much larger. Otherwise, the theory can't reconcile with all the experiments. Unfortunately, it is still unknown what determines the scale of $a$. In this paper, however, we found that  $a$ may be a universal constant. It is surprising that the value of $a$ we obtained here is the same as the upper bound value constrained from the Cassini spacecraft data \citep{Grumiller:2011gg}. Furthermore, a fitting to the Burkert's dark matter profile shows that within the scale length $r_0$, the mean dark matter surface density is approximately a constant for galaxies spanning a luminosity range of 14 magnitudes. This leads to a constant gravity acceleration at the radius $r_0$, i.e. $\mathbf{g}(r_0)\approx 0.3 \times 10^{-10}$ m s$^{-2}$ \citep{Donato2009,Gentile2009}, which is in good agreement with our result. It is interesting that the same acceleration is measured in so many systems. This motivates us to investigate the modified gravity theories in which such a scale exists naturally.

The most interesting feature of Grumiller's gravity is that it predicts a rotation curve proportional to the square root of distance in the large distance limits. However, the observed data just extend to a few tens of kpc. Although the velocity law Eq.(\ref{eq:fit-formula}) follows well to the experiment data in this range, it's too early to say that this is a valid model at large distances. Future observations of rotation velocity at large distances may provide further tests to the theory. It is very likely that Grumiller's gravity will fail in the large distance limits if $a$ is a constant. As was mentioned by \citet{Grumiller:2010bz}, $a$ may be a function of $r$. If we choose $a=\sqrt{GMa_0}/r$, Eq.(\ref{eq:Grumiller-velocity}) becomes
\begin{equation}\label{eq:mond-velocity}
v^2=v_N^2+\sqrt{GM(r)a_0},
\end{equation}
where $a_0\approx1.2\times 10^{-10}$ m s$^{-2}$ is the critical acceleration in Milgrom's MOND and $M(r)$ is the mass surrounded by a sphere of radius $r$. Eq.(\ref{eq:mond-velocity}) has the same asymptotic behavior as MOND, with the Tully-Fisher relation holds. \citet{Li2012} obtained a similar formula with the same asymptotic behavior as Eq.(\ref{eq:mond-velocity}) based on the Finsler geometry. \citet{Lietal2012} showed that Eq.(\ref{eq:mond-velocity}) plus the dipole and quadrupole contributions can well explain the mass discrepancy of Bullet Cluster 1E0657-558. Since the observed data show that most galaxies have asymptotically flat rotation curves, it may better describe the galaxy rotation curves if $a$ is inversely proportional to $r$ than is a constant.

\section*{Appendix}

\subsection*{A: Spherical Reduction}

For a spherical metric, it is possible to reduce the 4-dimensional Einstein-Hilbert action to a specific 2-dimensional dilaton gravity model. Generally, a 4-dimensional spherical spacetime manifold $M$ can be decomposed to the direct product of two 2-dimensional  sub-manifolds, $M=L\otimes S$, where the radial part $L$ depends only on the coordinates ($t,r$). At the same time, the Ricci scalar on $M$ induces the Ricci scalar on $L$. Given the 4-dimensional spherical metric
\begin{equation}\label{eq:line-element}
  ds^2=g_{\alpha\beta}dx^{\alpha}dx^{\beta}+\Phi^2(d\theta^2+\sin^2\theta d\phi^2),
\end{equation}
where the Greek index runs in ($t, r$) and the 2-dimensional metric $g_{\alpha\beta}$ and the dilaton field $\Phi$ depend only on ($t, r$), a straightforward calculation shows that \citep{Grumiller2001}
\begin{equation}\label{eq:RM-RL}
  R^{(M)}=R^{(L)}-\frac{2}{\Phi^2}[1+(\nabla\Phi)^2]-\frac{4}{\Phi}(\Delta\Phi),
\end{equation}
where  $R^{(M)}$ and $R^{(L)}$ are the Ricci scalar on $M$ and $L$, respectively.  `$\nabla$' and `$\Delta$' are the covariant derivative operator and the Laplacian operator on $L$, respectively.

On the other hand, the determinant of the metric on $M$ can be written as
\begin{equation}\label{eq:det-g}
  g^{(M)}=g^{(L)}\Phi^4\sin^2\theta.
\end{equation}
Substituting Eq.(\ref{eq:RM-RL}) and Eq.(\ref{eq:det-g}) into the Einstein-Hilbert action, and integrating over the angles ($\theta,\varphi$), we get
 \begin{equation}\label{eq:EH-action2}
   S_{\rm EH}=-\int d^4x\sqrt{-g^{(M)}}R=-4\pi \int d^2x\sqrt{-g^{(L)}}\left[\Phi^2R^{(L)}-2\left(1+(\nabla\Phi)^2\right)-4\Phi\Delta\Phi\right].
 \end{equation}
Integrating the last term by part and dropping the surface term, we get
\begin{equation}
  \int d^2x\sqrt{-g^{(L)}}(-4\Phi\Delta\Phi)=4\int d^2x\sqrt{-g^{(L)}}(\nabla\Phi)^2.
\end{equation}
 Thus, Eq.(\ref{eq:EH-action2}) becomes
 \begin{equation}
      S_{\rm EH}=-4\pi \int d^2x\sqrt{-g^{(L)}}\left[\Phi^2R^{(L)}+2(\nabla\Phi)^2-2\right].
 \end{equation}
Similarly, for a spherically symmetric matter Lagrangian $\mathcal{L}_{\rm m}=\mathcal{L}_{\rm m}(t,r)$, the 4-dimensional action can be reduced to a 2-dimensional one
\begin{equation}
  S_{\rm m}=-\int d^4x\sqrt{-g^{(M)}}\mathcal{L}_{\rm m}=-4\pi\int d^2 x\sqrt{-g^{(L)}}\Phi^2\mathcal{L}_{\rm m}.
\end{equation}

\subsection*{B: Ricci Scalar}

For a given 4-dimensional spherical metric
\begin{equation}
  g_{\mu\nu}={\rm diag}(-K^2,1/K^2,\Phi^2,\Phi^2\sin^2\theta),
\end{equation}
where $K^2\equiv1-\frac{2M}{r}-\Lambda r^2+2ar$ and $\Phi\equiv r$, the non-vanishing components of the metric and its inverse are
\begin{equation}
  g_{00}=-K^2,~~g_{11}=\frac{1}{K^2},~~g_{22}=\Phi^2,~~g_{33}=\Phi^2 \sin^2\theta,
\end{equation}
\begin{equation}
  g^{00}=-\frac{1}{K^2},~~g^{11}=K^2,~~g^{22}=\frac{1}{\Phi^2},~~g^{33}=\frac{1}{\Phi^2 \sin^2\theta}.
\end{equation}
A straightforward calculation shows that the non-vanishing Christoffel connections are given as
\begin{equation}
  \begin{dcases}
    \Gamma_{01}^{0}=\Gamma_{10}^{0}=\frac{1}{2}g^{00}g_{00,1}=\frac{1}{K}\frac{dK}{dr},\\
    \Gamma_{00}^{1}=-\frac{1}{2}g^{11}g_{00,1}=-K^3\frac{dK}{dr},\\
    \Gamma_{11}^{1}=\frac{1}{2}g^{11}g_{11,1}=-\frac{1}{K}\frac{dK}{dr},\\
    \Gamma_{22}^{1}=-\frac{1}{2}g^{11}g_{22,1}=-K^2\Phi\frac{d\Phi}{dr},\\
    \Gamma_{33}^{1}=-\frac{1}{2}g^{11}g_{33,1}=-K^2\Phi\frac{d\Phi}{dr}\sin^2\theta,\\
    \Gamma_{12}^{2}=\Gamma_{21}^{2}=\frac{1}{2}g^{22}g_{22,1}=\frac{1}{\Phi}\frac{d\Phi}{dr},\\
    \Gamma_{33}^{2}=-\frac{1}{2}g^{22}g_{33,2}=-\sin\theta\cos\theta,\\
    \Gamma_{13}^{3}=\Gamma_{31}^{3}=\frac{1}{2}g^{33}g_{33,1}=\frac{1}{\Phi}\frac{d\Phi}{dr},\\
    \Gamma_{23}^{3}=\Gamma_{32}^{3}=\frac{1}{2}g^{33}g_{33,2}=\cot\theta.
  \end{dcases}
\end{equation}
The non-vanishing components of the Ricci tensor are
\begin{equation}
  \begin{dcases}
    R_{00}=-K^2K'^{2}-K^3K''-\frac{2K^3}{\Phi}K'\Phi',\\
    R_{11}=-\frac{K''}{K}-\frac{2\Phi''}{\Phi}-\frac{2K'\Phi'}{K\Phi}-\frac{K'^2}{K^2},\\
    R_{22}=-2KK'\Phi\Phi'-K^2\Phi'^2-K^2\Phi\Phi''+1,\\
    R_{33}=(-2KK'\Phi\Phi'-K^2\Phi'^2-K^2\Phi\Phi''+1)\sin^2\theta.
  \end{dcases}
\end{equation}
The primes in the above equations denote the differentiation with respect to the radial distance $r$. Thus, the Ricci scalar is given as
\begin{equation}
  R\equiv R_{\mu\nu}g^{\mu\nu}=-\frac{2K^2\Phi''}{\Phi}+\frac{2}{\Phi^2}(-2KK'\Phi\Phi'-K^2\Phi'^2-K^2\Phi\Phi''+1).
\end{equation}

\section*{Acknowledgments}
We are grateful to Y. G. Jiang and S. Wang for useful discussions. This work has been funded in part by the National Natural Science Fund of China under Grant No. 10875129 and No. 11075166.

\newpage

\begin{table}
\begin{center}
\caption{Properties of the sample galaxies. Column (1): galaxy names. Columns (2) and (3): galaxy centers in J2000.0 from \citet{Walter2008}. Columns (4) and (5): Inclinations  and position angles from \citet{Walter2008}. Column (6): systematic velocities from \citet{Blok:2008}. Column (7): distances from \citet{Walter2008}. Column (8): the scale lengths of optical disk from \citet{Begeman1991} and \citet{Flores1993}, but corrected with the new distances in column (7). Column (9): mass of HI from \citet{Walter2008}. Column (10): apparent B-band magnitudes from \citet{Walter2008}. Column (11): absolute B-band magnitudes from \citet{Walter2008}. Column (12): luminosity calculated from column (11) using Eq.(\ref{eq:M-L-relation}).}
\begin{tabular}[t]{cccccccccccc}
\hline\hline
(1) & (2)& (3) & (4) & (5) & (6) & (7) & (8) & (9) & (10) & (11)  &  (12) \\
 Names   &   RA  &  DEC  &  INCL  &  PA  &  $V_{\rm sys}$ &  $D$   &  $h$ &  $M_{\rm HI}$  & $m_B$  &  $M_B$  &  $L$ \\
 &  [h m s]  &  [$^{\circ}~^{\prime}~^{\prime\prime}$] &  [$^{\circ}$]  &  [$^{\circ}$]  & [km/s] & [Mpc] &  [kpc]  &  $[10^8 M_{\odot}]$ & [mag]  & [mag]  & [$10^{10} L_{\odot}$] \\
 \hline
NGC2403 & 07 36 51.1 & +65 36 03 & 63 & 124 & 132.8 & 3.2 & 2.05 & 25.8 & 8.11 & -19.43 & 0.920\\
NGC2841 & 09 22 02.6 & +50 58 35 & 74 & 153 & 633.7 & 14.1 & 3.55 & 85.8 & 9.54 & -21.21 & 4.742\\
NGC2903 & 09 32 10.1 & +21 30 04 & 65 & 204 & 555.6 &  8.9 & 2.81  & 43.5  & 8.82 & -20.93 & 3.664\\
NGC3198 & 10 19 55.0 & +45 32 59 & 72 & 215 & 660.7 &  13.8 & 3.88 & 101.7 & 9.95 & -20.75 & 3.105\\
NGC3521 & 10 05 48.6 & -00 02 09 & 73 & 340 & 803.5 &  10.7 & 2.86 & 80.2 & 9.21 & -20.94 & 3.698\\
NGC5055 & 13 15 49.2 & +42 01 45 & 59 & 102 & 496.8 &  10.1 & 5.00 & 91.0 & 8.90 & -21.12 & 4.365\\
NGC7331 & 22 27 04.1 & +34 24 57 & 76 & 168 & 818.3 &  14.7 & 4.48 & 91.3 & 9.17 & -21.67 & 7.244\\
DDO154  & 12 54 05.9 & +27 09 10 & 66 & 230 & 375.9 &  4.3 & 0.54& 3.58 & 13.94 & -14.23 & 0.008\\
\hline
\end{tabular}\label{tab:samples}
\end{center}
\end{table}

\begin{table}
\begin{center}
\caption{The two-parameter best fit. The free parameters are $\Sigma_0$ and $a$. $M_{\rm disk}$ is the mass of the disk calculated from Eq.(\ref{eq:disk-mass}). $M/L$ is the mass-to-light ratio calculated from Eq.(\ref{eq:mass-to-light}). $\chi^2/n$ is the reduced chi-square. The numbers after ``$\pm$" are the $1 \sigma$ errors of the corresponding parameters.}
\begin{tabular}[t]{cccccc}
\hline\hline
& $\Sigma_0$ & $a$ & $M_{\rm disk}$ & $M/L$ & $\chi^2/n$\\
& $[M_{\odot}~\rm{pc}^{-2}]$ & $[10^{-10} \rm{m~s}^{-2}]$ & $[10^{10} M_{\odot}]$ & $[M_{\odot}/L_{\odot}]$ & \\
\hline
NGC2403 & $450.75\pm 9.73$   & $0.301\pm 0.006$ & $1.19\pm 0.03$ & $1.29\pm 0.03$ & 1.21\\
NGC2841 & $2626.12\pm 269.17$ & $0.396\pm 0.056$ & $20.79\pm 2.13$ & $4.39\pm 0.50$ & 2.47\\
NGC2903 & $1190.88\pm 24.66$ & $0.316\pm 0.010$ & $5.91\pm 0.12$ & $1.61\pm 0.03$ & 3.87\\
NGC3198 & $421.25\pm 9.75$  & $0.160\pm 0.005$ & $3.98\pm 0.09$ & $1.28\pm 0.03$ & 2.49\\
NGC3521 & $1239.17\pm 40.85$ & $0.296\pm 0.014$ & $6.37\pm 0.21$ & $1.72\pm 0.06$ & 0.54\\
NGC5055 & $900.27\pm 14.72$ & $0.053\pm 0.013$ & $14.14\pm 0.23$ & $3.24\pm 0.05$ & 5.84\\
NGC7331 & $1049.41\pm 53.30$ & $0.299\pm 0.021$ & $13.23\pm 0.67$ & $1.83\pm 0.09$ & 0.35\\
DDO154  & $92.76\pm 46.23$ & $0.060\pm 0.010$ & $0.02\pm 0.01$ & $2.12\pm 1.06$ & 17.48\\
\hline
\end{tabular}\label{tab:parameters1}
\end{center}
\end{table}

\begin{table}
\begin{center}
\caption{The one-parameter best fit. The free parameter is $\Sigma_0$. The parameter $a$ is  fixed at the average value $\bar{a}\approx0.30 \times 10^{-10} \rm{m~s}^{-2}$. $M_{\rm disk}$ is the mass of the disk calculated from Eq.(\ref{eq:disk-mass}). $M/L$ is the mass-to-light ratio calculated from Eq.(\ref{eq:mass-to-light}). $\chi^2/n$ is the reduced chi-square. The numbers after ``$\pm$" are the $1 \sigma$ errors of the corresponding parameters. We skip the one-parameter fit for NGC5055 and DDO154.}
\begin{tabular}[t]{ccccc}
\hline\hline
& $\Sigma_0$ & $M_{\rm disk}$ & $M/L$ & $\chi^2/n$\\
& $[M_{\odot}~\rm{pc}^{-2}]$  & $[10^{10} M_{\odot}]$ & $[M_{\odot}/L_{\odot}]$ & \\
\hline
NGC2403 & $444.96\pm 10.78$ & $1.17\pm 0.03$  & $1.28\pm 0.03$& 1.23\\
NGC2841 & $2825.29\pm 93.10$ & $22.37\pm 0.74$  & $4.72\pm 0.16$ & 4.15\\
NGC2903 & $1209.00\pm 34.15$  & $6.00\pm 0.17$  & $1.64\pm 0.05$ & 3.89\\
NGC3198 & $248.29\pm 13.20$ & $2.35\pm 0.12$  & $0.76\pm 0.04$ & 18.79\\
NGC3521 & $1222.09\pm 64.39$ & $6.28\pm 0.33$  & $1.70\pm 0.09$ & 0.55\\
NGC5055 & $\times$ &$\times$  & $\times$ & $\times$\\
NGC7331 & $1037.32\pm 72.80$ & $13.08\pm 0.92$ & $1.81\pm 0.13$ & 0.36\\
DDO154  & $\times$ & $\times$  & $\times$ & $\times$\\
\hline
\end{tabular}\label{tab:parameters2}
\end{center}
\end{table}

\begin{table}
\begin{center}
\caption{The three-parameter best fit. The free parameters are $\Sigma_0$, $a$ and $h$. $M_{\rm disk}$ is the mass of the disk calculated from Eq.(\ref{eq:disk-mass}). $M/L$ is the mass-to-light ratio calculated from Eq.(\ref{eq:mass-to-light}). $\chi^2/n$ is the reduced chi-square. The numbers after ``$\pm$" are the $1 \sigma$ errors of the corresponding parameters.}
\begin{tabular}[t]{ccccccc}
\hline\hline
& $\Sigma_0$ & $a$ & $h$ & $M_{\rm disk}$ & $M/L$ & $\chi^2/n$\\
& $[M_{\odot}~\rm{pc}^{-2}]$ & [$10^{-10}\rm{m~s}^{-2}$] & [kpc]  & $[10^{10} M_{\odot}]$& $[M_{\odot}/L_{\odot}]$ & \\
\hline
NGC5055& $1503.24\pm 217.05$ & $0.245\pm 0.042$ & $2.81\pm 0.32$ & $7.46\pm 2.78$ & $1.71\pm 0.64$ & 2.50\\
DDO154 & $47.86\pm 3.87$ & $0.027\pm 0.005$  & $2.45\pm 0.28$ & $0.18\pm 0.06$ & $22.56\pm 6.98$ & 0.86\\
\hline
\end{tabular}\label{tab:parameters3}
\end{center}
\end{table}

\label{lastpage}

\end{document}